\documentclass[debug]{rmaa}

\usepackage{paralist}

\usepackage{psfrag,color}

\usepackage[latin1]{inputenc}

\title{Radio Proper Motions and a Search for the Origin of PSR B1849+00} 

\author{
  L. F. Rodr\'iguez\altaffilmark{1,2},
  S. A. Dzib\altaffilmark{3},
  L. A. Zapata\altaffilmark{1}  and
  L. Loinard\altaffilmark{1} }

\altaffiltext{1}{Instituto de Radioastronom\'\i{}a y Astrof\'\i{}sica,
  UNAM, M\'exico.}

\altaffiltext{2}{Mesoamerican Center for Theoretical Physics, 
UNACH, 
M\'exico}

\altaffiltext{3}{Max-Planck-Institut fur Radioastronomie, Germany}

\shortauthor{Rodr\'iguez, Dzib, Zapata \& Loinard}
\shorttitle{Radio Proper Motions of PSR B1849+00}

\fulladdresses{
\item Sergio A. Dzib: Max-Planck-Institut f\"{u}r Radioastronomie, Auf dem H\"{u}gel 69, D-53121 Bonn, Germany. 
\item Laurent Loinard, Luis F. Rodr\'iguez \& Luis A. Zapata: Instituto de Radioastronom\'\i{}a y
  Astrof\'\i{}sica, Universidad Nacional Aut\'onoma de M\'exico,
  Apartado Postal 3--72, 58090 Morelia, Michoac\'an, M\'exico
  (l.rodriguez@irya.unam.mx).}

\listofauthors{L. F. Rodr\'iguez \&
  S. A. Dzib, L. A: Zapata \& L. Loinard}
\indexauthor{Rodr\'iguez, L. F.}
\indexauthor{Dzib, S. A.}
\indexauthor{Zapata, L. A.}
\indexauthor{Loinard, L.}

\abstract{Until now it has not been possible to obtain the proper motions of PSR B1849+00
with timing techniques or
VLBI imaging given the enhanced interstellar scattering along 
its line of sight. We present an analysis of archive Very Large Array
observations at epochs from 2012 to 2022 that indicates 
a total proper motion of 23.9$\pm$5.5 mas yr$^{-1}$ toward the southwest. After
correction for the proper motions produced by galactic rotation,
we find a peculiar transverse velocity of $\simeq$740 km s$^{-1}$.
We searched unsuccessfully along the past trajectory of 
the pulsar for an associated supernova remnant. In particular, W44 is in this trajectory but its distance is different to that of
PSR B1849+00.}

\resumen{Hasta ahora no ha sido posible obtener el movimiento propio de 
PSR B1849+00 con t\'ecnicas de sincronizaci\'on o im\'agenes de VLBI debido a la
fuerte dispersi\'on en su l\'inea de visi\'on.
Presentamos un an\'alisis de datos del archivo del Very Large Array
en \'epocas de 2012 a 2022 que indican un movimiento propio total de 23.9$\pm$5.5 mas yr$^{-1}$ 
hacia el suroeste. Despu\'es de corregir por el movimiento propio producido por la
rotaci\'on gal\'actica encontramos una velocidad transversal peculiar de $\simeq$740 km s$^{-1}$.
Buscamos infructuosamente en la trayectoria pasada del pulsar
por una remanente de supernova asociada. En particular, W44 est\'a en esta trayectoria pero
su distancia es diferente a la de 
PSR B1849+00.}

\addkeyword{Proper motions (1295)}
\addkeyword{Radio astrometry (1337)}
\addkeyword{Radio pulsars (1353)}

\begin{document}
\maketitle

\section{Introduction}
\label{sec:intro}

The positions and proper motions of hundreds of pulsars have been determined using the analysis of pulsar timing residuals
(e.g. Hobbs et al. 2004) and Very Long Baseline Interferometry (VLBI) imaging (e.g. Deller et al. 2019). However, 
in some cases, the pulsar suffers
strong scattering effects from plasma along the line of sight. This condition
scatters the radio image of the pulsar up to arcsec angular dimensions,
ruling out the possibility of VLBI imaging. For some sources it has been possible to make images with connected interferometers
(i.e. the Very Large Array), since these observations do not resolve out the scattered emission
(e.g. Dzib et al. 2018). Multiple examples of pulsars whose
proper motions have been determined with the Very Large Array are given by Brisken et al. (2003). 
The presence of inhomogeneous plasma in the
line of sight also limits the timing method by broadening the pulses to a level that makes timing experiments difficult or impossible.

PSR B1849+00 (PSR J1852+00) is a pulsar that has a period of 2.18 s and a characteristic age of $10^{5.55}$ yr
(Taylor et al. 1993). It is located in a region of enhanced line-of-sight interstellar scattering (Lazio 2004) that makes it
one of the most heavily scattered pulsar known, with a pulse
broadening time of 0.22 seconds at 1.4 GHz (L{\"o}hmer et al. 2001). It has a large dispersion measure,  DM = 787 cm$^{-3}$ pc 
(Han et al. 2016). Only seven of the 228 pulsars studied by these authors have larger dispersion measures.

In this paper we present an analysis of archive Very Large Array data with the goal of determining the proper motions of 
PSR B1849+00 and possibly advance in our understanding of the supernova that created this pulsar.

\section{VLA  Observations}
\label{sec:observations}

We searched unsuccessfully in the archives of the 
Karl G. Jansky VLA of NRAO\footnote{The National 
Radio Astronomy Observatory is a facility of the National Science Foundation operated
under cooperative agreement by Associated Universities, Inc.} 
for observations made in the highest angular resolution A configuration pointing toward PSR B1849+00.
We found, however, several good quality A configuration observations made 
in the L-band (1-2 GHz) with the phase center at the position of the gain
calibrator J1851+0035. Even when J1851+0035 is located at $\sim 11'$ from PSR B1849+00
we could obtain images of good quality over the full extent of the
primary beam ($\sim 30'$). This was possible because the Jansky VLA records the continuum data in
narrow channels (2 MHz), that do not produce significant bandwidth smearing. For all observations the position of J1851+0035
used was the updated position given in the VLA Calibrator Manual, RA(J2000) = $18^h 51^m 46\rlap.^s7217$; 
DEC(J2000) = $00^\circ 35' 32\rlap.{''}414$. The data were calibrated in the standard manner using the CASA (Common Astronomy Software Applications;  McMullin et al. 2007) package of NRAO and
the pipeline provided for VLA\footnote{https://science.nrao.edu/facilities/vla/data-processing/pipeline} observations. We made images using a robust weighting (Briggs 1995) of 0 to
optimize the compromise between angular resolution and sensitivity.  All images were also corrected for the primary beam
response. Finally, the images were also corrected for wide-field effects
using the gridding option  \it widefield \rm with 10$\times$10 subregions in the task TCLEAN (Rodriguez et al. 2023).

\begin{table*}[!t]\centering
  \scriptsize
 \newcommand{\DS}{\hspace{1\tabcolsep}} 
  \begin{changemargin}{-0.0cm}{-2cm}
    \caption{\hskip -2.0cm Parameters of the Jansky VLA observations}
    \setlength{\tabnotewidth}{0.84\linewidth}
    \setlength{\tabcolsep}{1.2\tabcolsep} \tablecols{7}
    \begin{tabular}{l @{\DS} ccccccc}
      \toprule
      &Mean & $\nu$ & Synthesized & Flux Density & \multicolumn{2}{c}{Position of PSR B1849+00}
       \\
      Project & Epoch & (GHz) & Beam & (mJy) &
      RA(J2000)\tabnotemark{a}  & DEC(J2000)\tabnotemark{b}   
      \\
      \midrule
      12B-225 & 2012.779 & 1.50 &  
      1$\rlap.{''}$47$\times$1$\rlap.{''}$21; +20$\rlap.^{\circ}$6  & 3.51$\pm$0.33 & 27$\rlap.^{s}$499$\pm$0$\rlap.^{s}$002 & 
      01$\rlap.{''}$91$\pm$0$\rlap.{''}$03  \\ 
     14A-404 & 2014.362 & 1.52 & 2$\rlap.{''}$25$\times$1$\rlap.{''}$60; $-$44$\rlap.^{\circ}$4 & 4.74$\pm$0.20 &
       27$\rlap.^{s}$502$\pm$0$\rlap.^{s}$002& 01$\rlap.{''}$83$\pm$0$\rlap.{''}$02  
       \\
       15A-301 & 2015.630 & 1.50 & 2$\rlap.{''}$05$\times$1$\rlap.{''}$35; $-$48$\rlap.^{\circ}$4 & 6.04$\pm$1.00&
       27$\rlap.^{s}$507$\pm$0$\rlap.^{s}$007 & 01$\rlap.{''}$80$\pm$0$\rlap.{''}$05 
        \\
       19A-386 & 2019.647 & 1.57 & 1$\rlap.{''}$39$\times$1$\rlap.{''}$21; +24$\rlap.^{\circ}$8 & 3.92$\pm$0.32 &
       27$\rlap.^{s}$494$\pm$0$\rlap.^{s}$004 & 01$\rlap.{''}$80$\pm$0$\rlap.{''}$03 
       \\
     22A-097 & 2022.474 & 1.52 & 2$\rlap.{''}$29$\times$1$\rlap.{''}$32; +65$\rlap.^{\circ}$3 & 2.42$\pm$0.21 &
       27$\rlap.^{s}$486$\pm$0$\rlap.^{s}$004 & 01$\rlap.{''}$72$\pm$0$\rlap.{''}$02 
       \\ 
                \bottomrule
       \tabnotetext{a}{\scriptsize Offset from RA(J2000) = $18^h 52^m 00^s$.} 
     \tabnotetext{b}{\scriptsize Offset from DEC(J2000) = $+00^\circ 32' 00''$.} 
    \end{tabular}
  \end{changemargin}
\end{table*}

\section{Discussion}
\label{sec:discussion}

In Table 1 we list the 
five projects found, indicating the name of the project, its mean epoch, frequency, synthesized beam, and flux density.
We finally give the position of PSR B1949+00 for each epoch. 
In Figure 1 we show a contour image of the source
from project 22A-097. In Figure 2 we present the positions as a function of time for  PSR B1949+00.
In Table 2 we give the equatorial
proper motions obtained from a least-squares fit to the positions shown in Figure 2.
In this Table we can see that 
PSR B1949+00 has a statistically significant equatorial total proper motion (4.3$-\sigma$).  

The weighted mean flux density and
weighted mean standard deviation at 1.5 GHz over the five epochs are 3.7$\pm$1.1 mJy for PSR B1949+00.
The spectral index can be used to further define the nature of the radio source. 
Kijak et al. (2011) report a spectral index of $\alpha = -2.4$ for this source, a steep
value characteristic of pulsars (e.g. Taylor et al. 1993; Maron et al. 2000).

\begin{figure*}[!t]
 \vskip-1.8cm
  \includegraphics[width=0.82\linewidth]{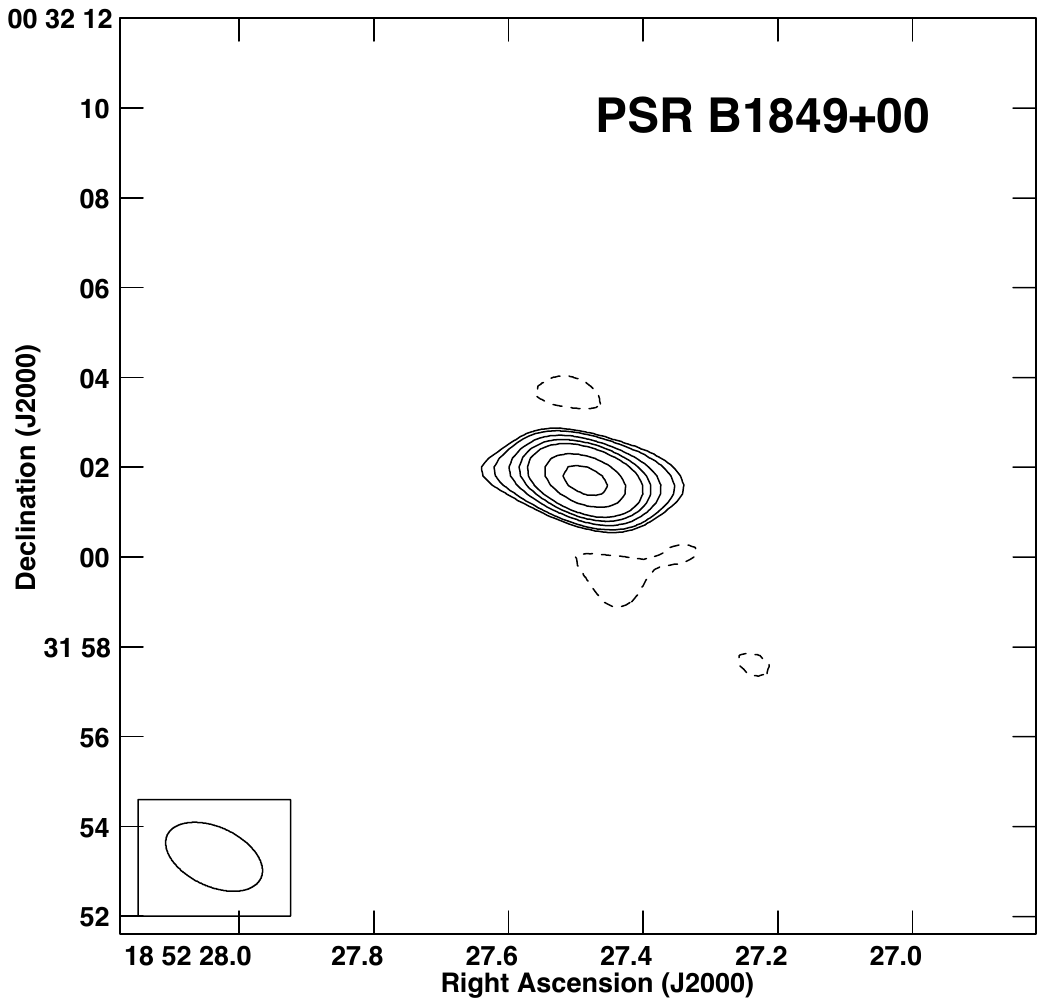}%
  \vskip-2.3cm
    \caption{\scriptsize Very Large Array contour image of PSR B1849+00 
    at 1.52 GHz for epoch 2022.474. Contours 
    are -3, 3, 4, 6, 8, 10, 15, and 20
    times 100  $\mu$Jy beam$^{-1}$, the rms noise in this region of the image. The synthesized beam 
    ($2 \rlap.{"}29 \times 1\rlap.{"}32; +65\rlap.^\circ3$)
    is shown in the bottom left corner of the image. 
    }
  \label{fig:widefig2}
\end{figure*}

\begin{table*}[!t]\centering
  \scriptsize
 \newcommand{\DS}{\hspace{1\tabcolsep}} 
  \begin{changemargin}{-2.8cm}{-2cm}
    \caption{Position and Proper Motions of PSR B1849+00}
    \setlength{\tabnotewidth}{0.85\linewidth}
    \setlength{\tabcolsep}{0.5\tabcolsep} \tablecols{10}
    \begin{tabular}{l @{\DS} cccccccccc}
      \toprule
      &\multicolumn{2}{c}{Position\tabnotemark{a} \label{tab:par} } &\multicolumn{4}{c}{Equatorial Proper Motions\tabnotemark{d}} 
     &\multicolumn{4}{c}{Corrected Galactic Proper Motions\tabnotemark{d}} \\ 
     & 
      RA(J2000)\tabnotemark{b} \label{tab:par} & DEC(J2000)\tabnotemark{c} \label{tab:par}  & $\mu_{RA}$cos(DEC)   & $\mu_{DEC}$ 
      & $\mu_{TOTAL}$ & PA\tabnotemark{e} 
    & $\mu_{l}cos(b) $  & $\mu_{b}$ 
     & $\mu_{TOTAL}$ & PA\tabnotemark{e}  \\
      \midrule
 & 27$\rlap.^{s}$518$\pm$0$\rlap.^{s}$007 & 02$\rlap.{''}$06$\pm$0$\rlap.{''}$06 & -19.1$\pm$6.4 & -14.4$\pm$3.2 &  23.9$\pm$5.5 & $233^\circ\pm9^\circ$ 
&  -14.6$\pm$5.8 & +10.4$\pm$4.2 &  17.9$\pm$5.3 & $305^\circ\pm15^\circ$\\  
           \bottomrule
      \tabnotetext{a}{\scriptsize For epoch 2000.0.}
             \tabnotetext{b}{\scriptsize Offset from RA(J2000) = $18^h 52^m 00^s$.} 
      \tabnotetext{c}{\scriptsize Offset from DEC(J2000) = $+00^\circ 32' 00''$.}
       \tabnotetext{d}{\scriptsize In mas yr$^{-1}$.}
       \tabnotetext{e}{\scriptsize Position angle of the proper motion vector in the respective coordinates.}       
    \end{tabular}
  \end{changemargin}
\end{table*}


\begin{figure*}
\vskip-0.2cm
\hskip-0.9cm
  \includegraphics[angle=0,width=1.20\linewidth]{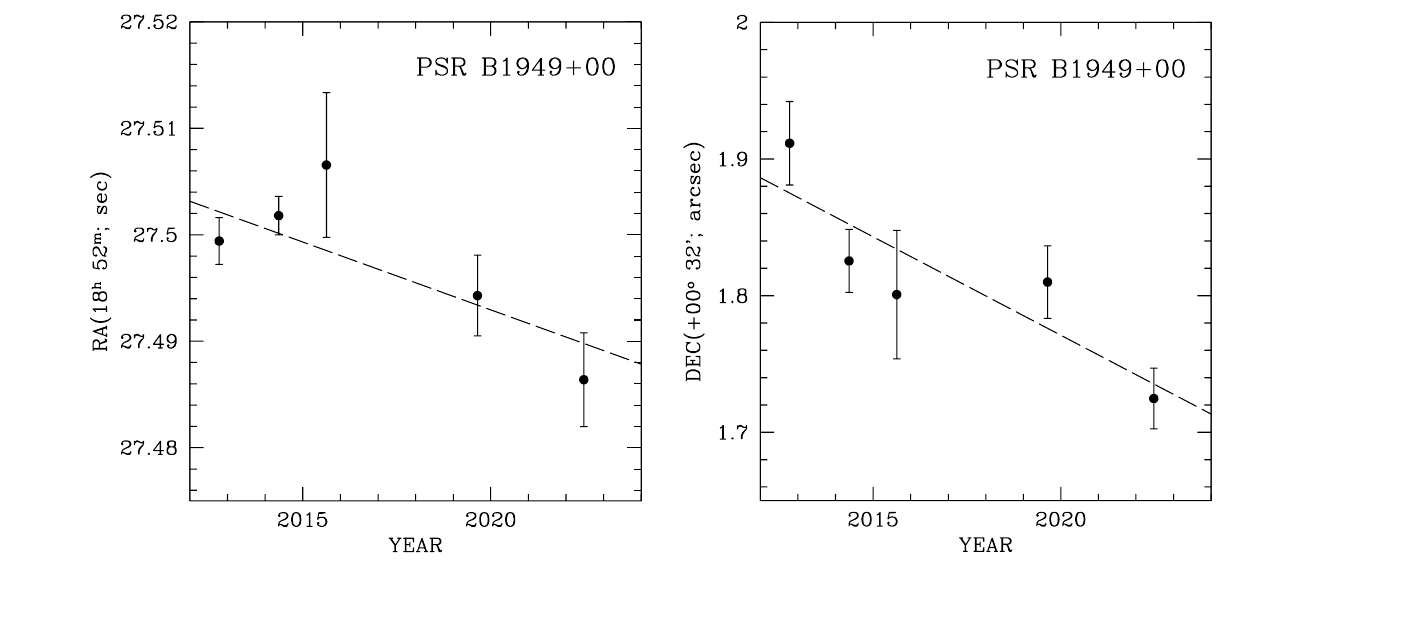}%
  \hfill
   \hskip-0.8cm
   \vskip-0.8cm
  \caption{\scriptsize (Left) Right ascension and (right) declination of PSR B1849+00 as a function of time.
  The dashed lines indicate the least squares fit for each parameter. The resulting proper motions are given in Table 2.}
  \label{fig:pm}
\end{figure*}

\section{The Transverse Velocity of PSR B1849+00}
\label{sec:velocity}

The distance of PSR B1849+00 has been estimated by several groups. We weight-averaged the results of Cordes \& Lazio (2003),
Verbiest et al. (2012), Yao et al. (2017), Pynzar (2020) and K{\"u}t{\"u}kc{\"u} et al. (2022), listed in Table 3,
to obtain a value of 8.6$\pm$1.7 kpc. 

The total proper motions of PSR B1949+00 have to be corrected for the proper motion contribution due to the
kinematics of the Galaxy in order to obtain the peculiar proper motions.
We use a distance from the Sun to the center of the Galaxy of 8.15 kpc and a circular rotation velocity at the Sun 
of 236 km s$^{-1}$ (Reid et al.
2019). We also assume a flat rotation curve outside the tangent point (with a galactocentric distance of 4.6 kpc
in the direction of the pulsar) and that the H I disk of the Galaxy
has an outer radius of 13.4 kpc (Goodwin, Gribbin,
\& Hendry 1998). In Figure 3 we show the 
proper motion in galactic longitude for a circular
Galactic orbit
as a function of distance in the direction of PSR B1949+00.
At the distance of the pulsar, 8.6 kpc, a galactic longitude proper motion of $-$6.9 mas yr$^{-1}$ is expected. We have corrected the 
galactic proper motions of PSR B1949+00 for this effect to obtain the values given in Table 2. This correction reduces the total
proper motion from 23.9 mas yr$^{-1}$ to 17.9 mas yr$^{-1}$. This is the peculiar proper motion of the pulsar.
\begin{figure*}[!t]
  \includegraphics[width=0.80\linewidth]{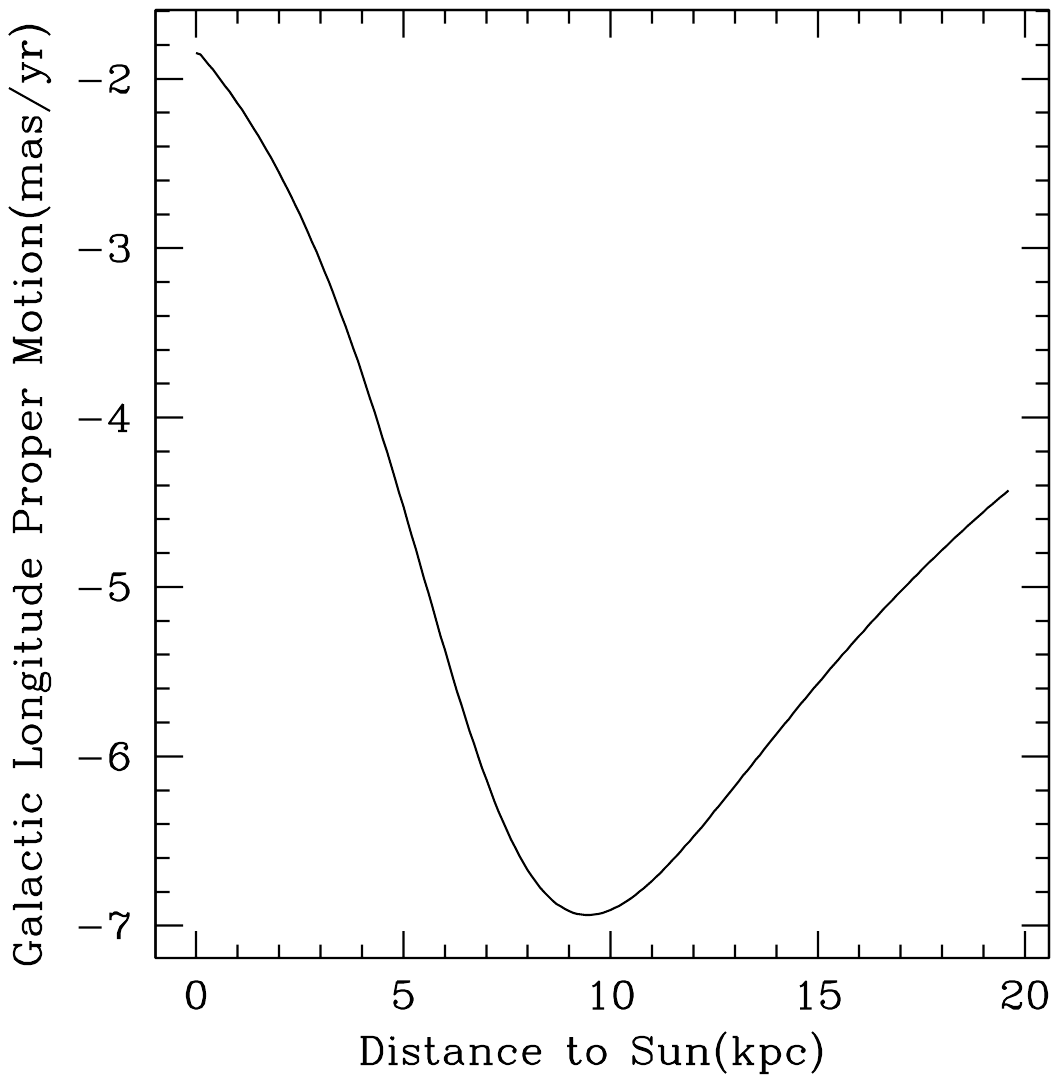}%
    \vskip-2.5cm
  \caption{\scriptsize Galactic longitude proper motion in the direction of PSRB1849+00 as a function of distance to the Sun.
This theoretical curve is based on the Galactic model
for circular motion described in the text.
 }
  \label{fig:pm}
\end{figure*}

At the distance of 8.6 kpc the corrected total proper motion implies a peculiar velocity of
740$\pm$220 km s$^{-1}$ in the plane of the sky. 
The large error comes from propagating the errors 
in the proper motion and in
the distance. 
The Australia Telescope National Facility Pulsar Catalogue (Manchester et al. 2005) lists a total of 294 pulsars
with reported transverse velocities. Of these pulsars, 15 (5\%) equal or exceed 740  km s$^{-1}$.

\begin{table*}[!t]\centering
\small
 \newcommand{\DS}{\hspace{1\tabcolsep}} 
  \begin{changemargin}{-0.0cm}{-2cm}
    \caption{\hskip -2.5cm Distance Estimates to PSR B1849+00}
    \setlength{\tabnotewidth}{0.94\linewidth}
    \setlength{\tabcolsep}{1.8\tabcolsep} \tablecols{3}
    \begin{tabular}{l @{\DS} ccc}
      \toprule
      Distance(kpc) & Method & Reference \\     
       \midrule
   8.4$\pm$1.7 & Dispersion measure model & Cordes \& Lazio (2003) \\
8.0$\pm$2.0 & HI absorption & Verbiest et al. (2012) \\
7.0$\pm$1.0 & HI absorption & Yao et al. (2017)  \\
 10.7$\pm$ 0.9 & Dispersion measure model & Pynzar et al. (2020) \\
7.0$\pm$1.5 & Dispersion measure model &  K{\"u}t{\"u}kc{\"u} et al. (2022) \\
                            \bottomrule
    \end{tabular}
  \end{changemargin}
\end{table*}



\section{The possible origin of PSR B1849+00}
\label{sec:origin}

In Figure 4 we plot the position and proper motion of PSR B1849+00, superposed on a GLOSTAR image at 5.8 GHz
(Medina et al. 2019;
Brunthaler et al. 2021; Medina et al. 2024).
In this image we also mark with circles supernova remnants and candidate supernova remnants in the
region.

The sources best aligned with the past trajectory of PSR B1849+00 are the W44 remnant and the supernova remnant candidate G34.524-0.761.
Taking the characteristic age of 10$^{5.55}$ yr and the corrected total proper motion given in Table 2, we find that assuming
a ballistic motion the origin would be located at 
$\sim1\rlap.^\circ77$ to the NE of its present position. The supernova W44 is located at $1\rlap.^\circ16$ to the NE of PSR B1849+00
and sounds like an interesting possibility since pulsar characteristic ages could depart importantly from the true age of the pulsar 
(Suzuki et al. 2021). 

However, the distance of W44 appears to be well established at  $\simeq$3 kpc (Radhakrishnan et al. 1972; 
Caswell et al. 1975; Cox et al. 1999;
Su et al. 2014; Wang et al. 2020). The first four groups used HI absorption observations and a model of the Galactic kinematics, 
while the last one used the extinction to red clump stars (Paczy{\'n}ski \& Stanek 1998)
probably associated with the supernova.
This significantly different distance seems to rule out an association of W44 with PSR B1849+00. In any case, it would be valuable to
estimate the distance to W44 doing parallax measurements of the OH 1720 MHz masers associated with it.
Also W44, whose age is estimated to be $\simeq$20,000 yr (Smith et al. 1985;
Cox et al. 1999;  Giuliani et al. 2011), seems to be much younger that the pulsar and this result also gravitates 
against an association. The supernova age is estimated 
from observations of its size, expansion rate, and the properties of the surrounding interstellar medium.

\begin{figure*}[!t]
 \vskip-0.0cm
  \includegraphics[width=1.1\linewidth]{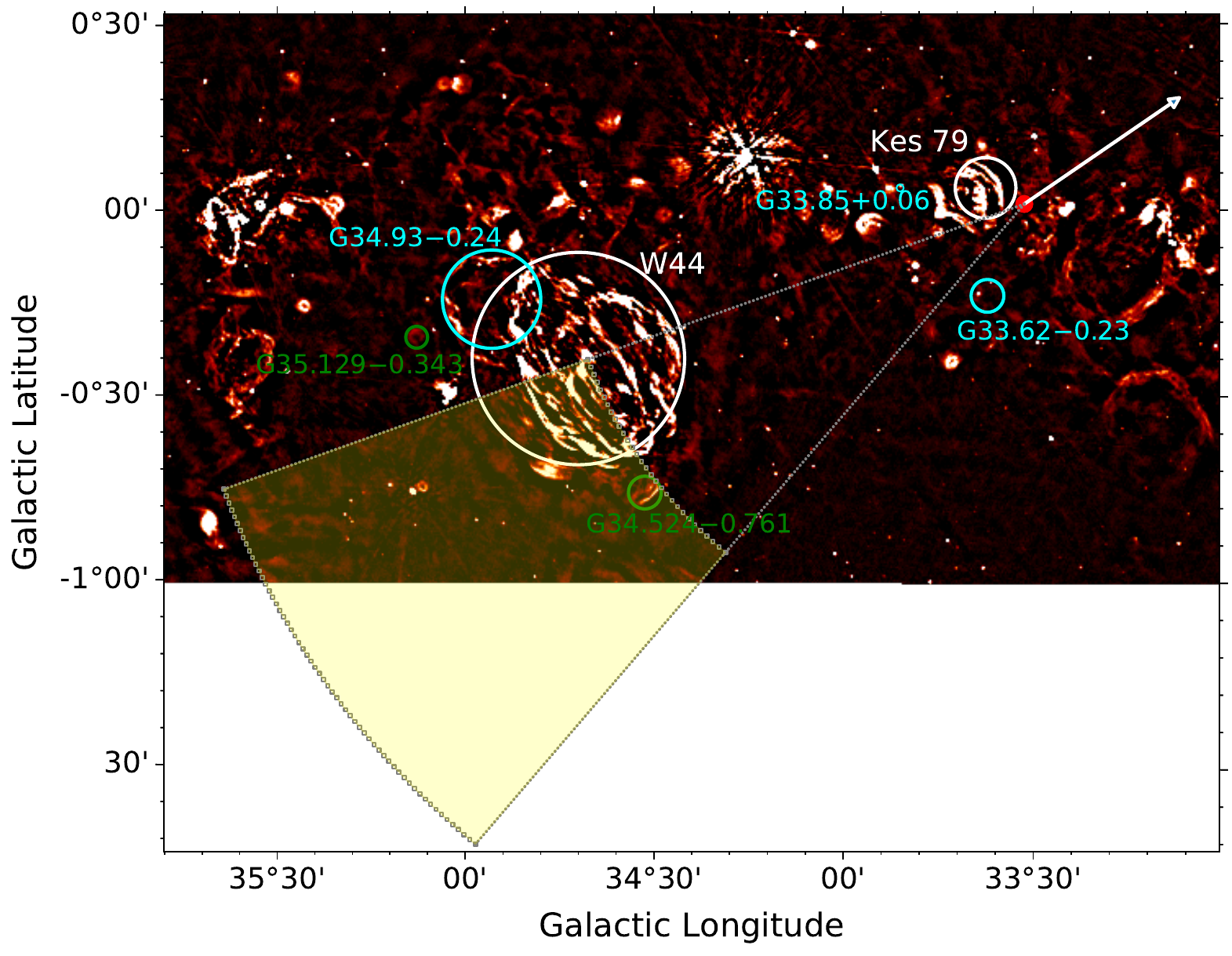}%
  \vskip-0.0cm
    \caption{\scriptsize Position (red dot) and proper motion of PSR B1849+00 for a period of $10^5$ yr (white arrow)
in galactic coordinates and    
superposed on a GLOSTAR image. The dashed lines indicate the $\pm$1-$\sigma$ angular range for the past trajectory of the pulsar. The dashed curved lines indicate the $\pm$1-$\sigma$ distance range for the
origin of the pulsar assuming an age of 10$^{5.55}$ yr,
the characteristic age of the pulsar. These dashed lines define the most likely area in the sky for the origin of the pulsar, shown in yellow.
The white circles indicate the position and extent of confirmed
supernova remnants (Green 2022), while the green and cyan circles indicate candidate supernova remnants from Dokara et al. (2021)
and Anderson et al. (2017), respectively.
 }
  \label{fig:pm}
\end{figure*}

The distance to G34.524-0.761 is not yet determined. This source has a spectral index of -0.9 and exhibits a 10\% degree of linear polarization (Dokara et al. 2023), supporting the supernova remnant interpretation. However, as noted by these authors, 
its filamentary morphology suggests that it is probably a fragment of a much larger faint remnant. We conclude that
we cannot associate clearly a supernova remnant with PSR B1849+00. It is probable that the related supernova remnant
has mixed with the interstellar medium and is no longer detectable.

The Next Generation VLA (Murphy et al. 2018), with its unprecedented sensitivity and high angular resolution
will be the ideal instrument for the study of the proper motions of pulsars whose image and pulses are scattered
by inhomogeneous plasma in the line of sight.

\section{Conclusions}

1) We analyzed archive VLA observations of the pulsar PSR B1849+00
to obtain its radio proper motions. At a distance of 8.6 kpc the resulting peculiar velocity in the plane of the sky is 
$\simeq$740 km s$^{-1}$. 

2) We searched unsuccessfully along the past trajectory of PSR B1849+00 for a supernova remnant 
that could be associated with this pulsar. The bright supernova remnant W44 is in the past trajectory 
of the pulsar, but its distance is different.We suggest that the remnant that originated PSR B1849+00 
has dissipated and is no longer detectable.

\begin{acknowledgments}


L.A.Z. acknowledges financial support from CONACyT-280775 and UNAM-PAPIIT IN110618, 
and IN112323 grants, M\'exico. L.L. acknowledges the support of DGAPA PAPIIT grant IN108324. L.F.R. 
acknowledges the financial support of DGAPA (UNAM) 
IN105617, IN101418, IN110618 and IN112417 and CONACyT 238631 and 280775-CF grant 263356. 
S.A.D. acknowledges the M2FINDERS project from the European Research
Council (ERC) under the European Union's Horizon 2020 research and
innovation programme (grant No 101018682).

\end{acknowledgments}

\end{document}